\documentstyle[12pt,dmtrieste]{article}
\def\mincir{\raise -2.truept\hbox{\rlap{\hbox{$\sim$}}\raise5.truept
\hbox{$<$}\ }}
\def\magcir{\raise -2.truept\hbox{\rlap{\hbox{$\sim$}}\raise5.truept
\hbox{$>$}\ }}

\catcode`\@=11
\long\def\@makefntext#1{
\protect\noindent \hbox to 3.2pt {\hskip-.9pt
$^{{\ninerm\@thefnmark}}$\hfil}#1\hfill}                

\def\@makefnmark{\hbox to 0pt{$^{\@thefnmark}$\hss}}  

\def\ps@myheadings{\let\@mkboth\@gobbletwo
\def\@oddhead{\hbox{}
\rightmark\hfil\ninerm\thepage}
\def\@oddfoot{}\def\@evenhead{\ninerm\thepage\hfil
\leftmark\hbox{}}\def\@evenfoot{}
\def\sectionmark##1{}\def\subsectionmark##1{}}
\def\ut#1{\mathop{\vtop{\ialign{##\crcr
     $\hfil\displaystyle{#1}\hfil$\crcr\noalign
     {\kern1pt\nointerlineskip}\hbox{$\hfil\sim\hfil$}\crcr
     \noalign{\kern1pt}}}}}

\def\undersymbol#1#2{\mathop{\vtop{\ialign{##\crcr
     $\hfil\displaystyle{#2}\hfil$\crcr\noalign
     {\kern1pt\nointerlineskip}\hbox{$\hfil#1\hfil$}\crcr
     \noalign{\kern1pt}}}}}

\begin{document}
\begin{flushright}
Lecce University LE-ASTRO 3/98\\
Pavia University FNT/T-98/01\\
Zurich University ZU-TH 7/98\\
\end{flushright}
\vspace*{2cm}
\centerline{\normalsize\bf MACHOs AS BROWN DWARFS\footnote{Talk given 
by M. Roncadelli at the workshop 
``DM-ITALIA-97'', Trieste, Italy (to appear in the proceedings).}}

\vspace*{0.6cm}
\centerline{\footnotesize FRANCESCO DE PAOLIS}
\centerline{\footnotesize\it Bartol Research Institute, University of Delaware}
\baselineskip=13pt
\vspace*{0.3cm}
\centerline{\footnotesize GABRIELE INGROSSO}
\centerline{\footnotesize\it Dipartimento di Fisica, Universit\`a di Lecce and 
INFN}
\baselineskip=13pt
\vspace*{0.3cm}
\centerline{\footnotesize PHILIPPE JETZER}
\centerline{\footnotesize\it Paul Scherrer Institut, Laboratory for 
Astrophysics}
\centerline{\footnotesize\it and Institute of Theoretical Physics, University 
of Zurich}
\vspace*{0.3cm}
\centerline{\footnotesize and}
\vspace*{0.3cm}
\centerline{\footnotesize MARCO RONCADELLI}
\baselineskip=13pt
\centerline{\footnotesize\it INFN, Sezione di Pavia}

\vspace*{0.9cm}
\abstracts{Recent observations of microlensing events in the Large Magellanic Cloud
suggest that a sizable fraction of the galactic halo is in the form of
Massive Astrophysical Compact Halo Objects (MACHOs). Although the average 
MACHO mass is presently poorly known, the value $\sim 0.1 M_{\odot}$ looks as 
a realistic estimate, thereby implying that brown dwarfs are a viable and 
natural candidate for MACHOs. We describe a scenario in which dark clusters of MACHOs and 
cold molecular clouds (mainly of $H_2$) naturally form in the halo at 
galactocentric distances larger than 10-20 kpc. Moreover, we discuss various 
experimental tests of this picture.}

\normalsize\baselineskip=15pt
\section{Introduction}

Since 1993 several microlensing events have been detected towards the Large 
Magellanic Cloud by the MACHO and EROS collaborations. Everybody agrees that 
this means that Massive Astrophysical Compact Halo Objects (MACHOs) have been 
discovered. Yet, the specific nature of MACHOs is unknown, mainly because 
their average mass turns out to depend strongly on the assumed galactic model. 
For instance, the spherical isothermal model would give $\sim 0.5 M_{\odot}$ 
whereas the maximal disk model would yield $\sim 0.1 M_{\odot}$ for that 
quantity. What can be reliably concluded today is only that MACHOs should lie 
in the mass range $0.05 M_{\odot} - 1~M_{\odot}$. Remarkably enough, the 
MACHO team has claimed that the fraction of galactic matter in the form of 
MACHOs is fairly model independent and -- within the present statistics -- 
should be $\sim 50 \%$. 

What is the most realistic galactic model? Regretfully, no clear-cut answer 
is presently available. Nevertheless, the current wishdom -- that the Galaxy 
ought to be best described by the spherical isothermal model -- seems less 
convincing than before and 
nowadays various arguments strongly favour a nonstandard galactic halo.
Indeed, besides the observational evidence that spiral galaxies generally have 
flattened halos, recent determinations of both the disk scale length, and the 
magnitude and slope of the rotation at the solar position indicate that
our galaxy is best described by the maximal disk model.
This conclusion is further strengthened 
by the microlensing results towards the galactic centre, which 
imply that the bulge is more massive than previously thought.
Correspondingly, the halo plays a less dominant r\^ole 
than within the spherical isothermal model, thereby reducing the halo microlensing 
rate as well as the average MACHO mass. A similar result occurs within
the King-Michie halo models, which also take
into account the finite escape velocity and the anisotropies in velocity 
space (typically arising during the phase of halo formation). Moreover, 
practically the same conclusions also hold  for flattened galactic models 
with a substantial degree of halo rotation. So, the expected average MACHO 
mass should be smaller than within the spherical isothermal model and the 
value $\sim 0.1~M_{\odot}$ looks as a realistic estimate to date.
This fact is of paramount importance, since it implies that brown dwarfs 
are a viable and natural candidate for MACHOs. Still -- even if MACHOs are 
indeed brown dwarfs --  the problem remains to explain their  
formation, as well as the nature of the remaining dark matter in 
galactic halos.

We have proposed a scenario in which dark clusters of MACHOs and 
cold molecular clouds -- mainly of $H_2$ -- naturally form in the halo at 
galactocentric distances larger than $10-20$ kpc (somewhat 
similar ideas have also 
been put forward by  Ashman and by Gerhard and Silk). 
Below, we shall review the main 
features of this model, along with its observational implications.

\section{Formation of dark clusters}
Our scenario encompasses the one originally proposed by Fall and Rees to 
explain the origin of globular clusters and can be summarized as follows.
After its initial collapse, the proto galaxy (PG) is expected to be shock
heated to its virial temperature $\sim 10^6$ K. Because of thermal 
instability, density enhancements rapidly grow as the gas cools. 
Actually, overdense regions cool
more rapidly than average, and so proto globular cluster
(PGC) clouds form in pressure equilibrium with hot diffuse gas. When the 
PGC cloud temperature reaches $\sim 10^4$ K, hydrogen recombination occurs: 
at this stage, their mass and size are
$\sim 10^5 (R/{\rm kpc})^{1/2} M_{~\odot}$ and $\sim 10~(R/kpc)^{1/2}$ pc,
respectively ($R$ being the galactocentric distance). Below $10^4$ K, the 
main coolants are $H_2$ molecules and any 
heavy element produced in a first chaotic galactic phase. As we shall see in 
a moment, the subsequent evolution of the PGC clouds will be very different 
in the inner and outer part of the Galaxy, depending on the decreasing 
ultraviolet (UV) flux as the galactocentric distance $R$ increases.

As is well known, in the central region of the Galaxy
an Active Galactic Nucleus (AGN) and a first population of  
massive stars are expected to form, which act as strong sources of UV 
radiation that dissociates the $H_2$ molecules. It is not difficult to 
estimate that $H_2$ depletion should happen for galactocentric distances 
smaller than $10-20$ kpc. As a consequence,
cooling is heavily suppressed in the inner halo, and so the PGC clouds here
remain for a long time at temperature $\sim 10^4$ K, resulting in the imprinting
of a characteristic mass $\sim 10^6 M_{\odot}$. Eventually, the UV flux will
decrease, thereby permitting the formation of $H_2$.
As a result, the cloud temperature drops below $\sim 10^4$ K and the subsequent 
evolution leads to star formation and ultimately to globular clusters.
 
Our main point is that in the outer halo -- namely for galactocentric 
distances larger than $10-20$ kpc -- no substantial $H_2$ depletion should 
take place (owing to the distance suppression of the UV flux).
Therefore, the PGC clouds cool and contract. When their number density 
exceeds $\sim 10^8$ cm$^{-3}$, further $H_2$ is produced via 
three-body reactions
($H+H+H \rightarrow H_2+H$ and $H+H+H_2 \rightarrow   2H_2$), which makes 
in turn the cooling efficiency increase dramatically.
This fact has three distinct implications:
(i) no imprinting of a characteristic PGC cloud mass shows up,
(ii) the Jeans mass can drop to values considerably smaller
than $\sim 1~M_{\odot}$, and
(iii)  the cooling time is much shorter than the free-fall time.
As pointed out by Palla, Salpeter and Stahler, in such a situation a 
subsequent fragmentation occurs into smaller and smaller clouds that remain 
optically thin to their own radiation. The process stops when the clouds 
become optically thick to their own line emission -- this happens when the
Jeans mass gets as low as $\sim 10^{-2}~M_{\odot}$. In this manner, dark
clusters should form, which contain brown dwarfs in the mass range 
$10^{-2}-10^{-1}~M_{\odot}$.

Before proceeding further, two observations are in order. First, it seems 
quite natural to suppose that -- much in the same way as it occurs for 
ordinary stars -- also in this case the fragmentation process that gives 
rise to individual brown dwarfs should produce a substantial fraction of 
binary brown dwarfs (they will be referred to as {\it primordial} binaries)
. It is important to keep in mind that the mass fraction of primordial 
binaries can be as large as $50\%$. Hence, we see that MACHOs consist of 
both individual and binary brown dwarfs in the present scenario. Second,
we do not expect the fragmentation process to be able to convert the whole 
gas in a PGC cloud into brown dwarfs. For instance, standard stellar formation 
mechanisms lead to an upper limit  of at most $40\%$ for 
the conversion efficiency. Thus, a substantial fraction $f$ of the primordial 
gas -- which is mostly $H_2$ -- should be left over. Because brown dwarfs 
do not give rise to stellar winds,
this gas should remain confined within a dark 
cluster. So, also cold $H_2$ self-gravitating clouds should presumably be 
clumped into dark clusters, along with some residual diffuse gas (the 
amount of diffuse gas inside a dark cluster has to be low, for otherwise it 
would have been observed in optical and radio bands).

Unfortunately, the total lack of any observational information about
dark clusters would make any effort to understand their structure and
dynamics practically impossible, were it not for some remarkable
insights that our unified treatment of globular and dark clusters provides 
us. In the first place, it looks quite natural to assume that also dark
clusters have a denser core surrounded by an extended spherical
halo. Moreover, in the lack of any further information it seems reasonable
to suppose (at least tentatively) that the dark clusters have the same
average mass density as globular clusters. Hence, we obtain
$r_{DC}\simeq 0.12~ ({M_{DC}}/{M_{\odot}})^{1/3}$  pc,
where $M_{DC}$ and $r_{DC}$ denote the mass and the median radius of a
dark cluster, respectively.
As a further implication of the above scenario, we stress that -- at
variance with the case of globular clusters -- the initial mass
function of the dark clusters should be smooth, since the monotonic
decrease of the PGC cloud temperature fails to single out any
particular mass scale. Finally, we suppose for definiteness 
(and with an eye to microlensing experiments) that all brown dwarfs have 
mass $\simeq 0.1~M_{\odot}$, while the molecular cloud spectrum will be 
taken to be $10^{-3}~M_{\odot}\ut < M_m \ut < 10^{-1}~M_{\odot}$.

\section{Dynamics of dark clusters}
As we have seen, MACHOs are clumped into dark clusters when they form in 
the outer galactic halo. Still, the further fate of these clusters is quite 
unclear.
For, they might either evaporate or drift towards 
the galactic centre. In the latter case, encounters with globular clusters 
might have dramatic observational consequences and dynamical friction 
could drive too many MACHOs into the galactic bulge. So, even if dark clusters 
are unseen, nontrivial constraints on their characteristic parameters arise 
from the observed properties of our galaxy. Moreover, in order to play any 
r\^ole as a candidate for dark matter, MACHOs must have survived until 
the present in the outer part of the galactic halo. Finally, it is important 
to know whether MACHOs are still clumped into clusters today, especially 
because an improvement in the statistics of microlensing observations permits 
to test this possibility.

Below, we shall schematically address various effects which concern the 
dynamics of dark clusters.

{\bf Dynamical friction} -- Dark clusters are subject to dynamical friction
as they orbit through the Galaxy, which makes them loose energy and therefore 
spiral in toward the galactic centre. It is straightforward to see that in our 
model -- since $R> 10-20 \ kpc$ and $M_{DC}\ut < 10^6~M_{\odot}$ -- a  
dark cluster originally at galactocentric distance $R$ will be closer to 
the galactic centre today by an amount $\Delta R\ut < 5.8\times 10^{-2}$ kpc. 
Therefore, dark clusters are still confined in the outer galactic halo.
As a consequence, encounters between dark and globular clusters as well as
disk and bulge shocking of dark clusters are dynamically irrelevant.

{\bf Encounters between dark clusters} -- Encounters between dark clusters may 
(under the circumstances to be analyzed below) lead to their disruption. 
What is important to notice is that the (one-dimensional) velocity dispersion
of dark clusters in the halo $\sigma\simeq 155$ km s$^{-1}$ is much larger 
than the (one-dimensional) velocity dispersion of MACHOs and molecular 
clouds inside a dark cluster 
$\sigma_*\simeq 7\times 10^{-2}({M_{DC}}/{M_{\odot}})^{1/3}$ km s$^{-1}$.
Hence we shall work within the impulse approximation. Our strategy is as 
follows. If we denote by $\Delta E$ the change of internal energy of a dark 
cluster in a single encounter, it is possible to express the rate $\dot{E}(R)$ 
at which the cluster's energy changes because of encounters in terms of 
$\Delta E$.
At this point, a natural definition of the time required by encounters to 
dissolve a cluster is provided by $t_d(R)=E_{\rm bind}/\dot{E}(R)$,
with $E_{\rm bind}\simeq 0.2~ GM_{DC}^2/r_{DC}$.
Demanding next that $t_d(R)$ should exceed the age of the Universe, we 
find under what conditions encounters will be harmless. It turns out 
that in our picture distant encounters are always harmless, whereas close 
encounters are harmless provided $M_{DC}\ut < 10^6~M_{\odot}$. 

{\bf Evaporation} -- As is well known, any stellar association evaporates 
within a finite time. Specifically, relaxation via gravitational two-body 
encounters leads to the escape of MACHOs approaching the unbound tail of 
the cluster velocity distribution, and this process gets enhanced by the 
tidal truncation of dark clusters due to the galactic gravitational field.
A key-r\^ole in the present analysis is played by the relaxation time. 
However, this is a local quantity, which can vary by various orders of 
magnitude in different regions of a single dark cluster.
Therefore it is more convenient to 
characterize a dark cluster by a single value of the relaxation time.
This goal is achieved by the introduction of the median relaxation
time $t_{\rm rh}$. Now, it turns out that the cumulative effect of several weak encounters -- 
which gradually increase the velocity of a given MACHO until it exceeds the 
local escape velocity -- dominates over single close encounters. 
Correspondingly, the evaporation time is $t_{\rm evap}\simeq 300~ t_{\rm rh}$. 
Therefore, by requiring that $t_{\rm evap}$ should exceed the age of the 
Universe we find that dark clusters with $M_{DC}\ut > 3\times 10^2
~M_{\odot}$ are not yet evaporated. It goes without saying that dark 
clusters are tidally disrupted by the galactic gravitational field unless 
$r_{DC}$ is smaller than their tidal radius. It is easy to see that this is 
always the case for $R>10-20$ kpc, and so within the present model dark 
clusters are not tidally disrupted by the galactic gravitational field.

{\bf Core collapse} -- Much in the same way as it happens for globular 
clusters, core collapse is expected to occur also for dark clusters. A 
thorough analysis shows that the dark clusters with $M_{DC}\ut <5 \times
10^4~M_{\odot}$ should have already began core collapse. Although the 
further fate of such clusters  crucially depends on the (unknown) model 
which describes them correctly, two points seem to be firmly established.
First, evaporation and subsequent mass ejection make the number of MACHOs 
in the dark clusters monotonically decrease with time. Second, in spite of 
the fact that the rise of central density leads to the formation of {\it 
tidally-captured} binary brown dwarfs in the cluster cores, their fraction 
turns out to be too small to play any r\^ole in the considerations to 
follow (however they are likely to stop and reverse core collapse).

\section{MACHOs as binary brown dwarfs}
As already pointed out, it seems natural to suppose that a fraction of 
primordial binary brown dwarfs -- possibly as large as $50\%$ in mass -- 
should form along with individual brown dwarfs as a result of the 
fragmentation process of the PGC clouds.
Subsequently, primordial binaries will concentrate inside the
core because of the mass stratification instability.
We recall that a binary system is hard when its internal energy
exceeds the kinetic energy of field stars. 
In the present case, binary brown dwarfs
happen to be hard when their orbital radius $a$ obeys the constraint 
$a<1.4\times 10^{12}({M_{\odot}}/{M_{DC}})^{2/3}~{\rm km}$. 
Now, consistency with the results of microlensing experiments demands that 
the overwhelming majority of binary brown dwarfs should be detected as 
unresolved objects. This requirement entails in turn that their orbital radius 
should be (roughly) less than one-half of the corresponding Einstein
radius, thereby implying that
the stronger bound $a<3\times 10^{8}~{\rm km}$ has to be satisfied.
As is well known, soft binaries always get softer whereas hard binaries always 
get harder, and so only those which 
are hard can survive until the present. Still, all values for the orbital 
radius of primordial binaries consistent with the above constraint are in 
principle allowed (tidally-captured binaries have $a\simeq 2.5\times
10^5$ km, and so they are no problem). Therefore, it is crucial to see 
whether any mechanism exists which makes primordial binaries  shrink 
in such a way that also the latter constraint is eventually obeyed.
One might think that collisional hardening -- namely the process whereby 
hard binaries get harder in encounters with individual brown dwarfs -- is 
able to do the job. However,  a detailed analysis shows that this is not 
the case. Remarkably enough, the goal is achieved by frictional hardening 
on molecular clouds (inside the dark cluster cores). That is, whenever a 
primordial binary crosses a cloud, dynamical friction operates. As a 
consequence, the binary releases binding energy, thereby getting harder. It 
can be seen that this process is very effective and indeed makes the 
overwhelming majority of primordial binary brown dwarfs unresolvable in 
microlensing experiments.

\section{Observational tests}
We list schematically some observational tests for the present scenario.

{\bf Clustering of microlensing events} -- The most promising way to detect 
dark clusters is via correlation effects in microlensing observations, as 
they are expected to exhibit a cluster-like distribution. Indeed, it has 
been shown that a relatively small number of microlensing events would be 
sufficient to rule out this possibility, while to confirm it more events 
are needed. However, we  have seen that core collapse can liberate a 
considerable fraction of MACHOs from the less massive clusters, and so an 
unclustered MACHO population is expected to coexist with dark clusters in 
the outer halo -- detection of unclustered MACHOs would therefore not 
disprove the present model.

{\bf $\gamma$-rays from halo clouds} -- A signature for the presence of 
molecular clouds in the galactic halo should be a $\gamma$-ray flux 
produced in the scattering of high-energy cosmic-ray protons on $H_2$.
As a matter of fact, an essential
ingredient is the knowledge of the cosmic ray flux in the halo. Unfortunately,
this quantity is unknown and the only available 
information comes from theoretical estimates.
More precisely,
from the mass-loss rate of a typical galaxy, we infer a total cosmic ray flux
in the halo $F \simeq 3.5\times 10^{-5}$ erg cm$^{-2}$ s$^{-1}$. We further
assume the same energy distribution of the cosmic rays as measured on Earth and
we rescale the overall density in such a way that the integrated energy 
flux agrees with the above value. Moreover, we suppose that the cosmic-ray 
density scales with $R$ like the dark matter density (i.e. $\sim R^{-2}$ in 
the outer halo).The best chance to detect the $\gamma$-rays in question is provided
by observations at high galactic latitude.
Accordingly, we find a $\gamma$-ray flux (for $E_{\gamma}>100$ MeV)
$\Phi_{\gamma}(90^0) \simeq ~(0.4 -1.8) \times 10^{-5}~f$ photons cm$^{-2}$
s$^{-1}$  sr$^{-1}$ if the cosmic rays are confined within the galactic halo, 
whereas we get $\Phi_{\gamma}(90^0) \simeq ~(0.6 -3) \times 10^{-7}~f$ photons 
cm$^{-2}$s$^{-1}$  sr$^{-1}$ if they are confined within the local galaxy 
group. These values should be compared with the experimental result from EGRET 
satellite $\Phi_{\gamma}(90^0) \simeq ~1.5 \times 10^{-5}$ photons cm$^{-2}$
s$^{-1}$  sr$^{-1}$. Actually, D. Dixon has re-analyzed the EGRET data and 
claims that the $\gamma$-ray flux from the halo is $\Phi_{\gamma}(90^0) 
\simeq ~2 \times 10^{-6}$ photons cm$^{-2}$s$^{-1}$  sr$^{-1}$. Thus, our 
values turn out to be neither too large nor too small to be interesting, and 
the $\gamma$-ray flux predicted by the present scenario may have already 
been detected! 
 
{\bf CBR anisotropy} -- An alternative way to discover the molecular 
clouds under consideration relies upon their emission in the microwave band. 
The temperature of the clouds has to be close to that of the cosmic background
radiation (CBR). Indeed, an upper limit of $\Delta T/T \sim 10^{-3}$ can
be derived by considering the anisotropy they would
introduce in the CBR due to their higher temperature. Realistically, molecular
clouds cannot be regarded as black body emitters because they mainly 
produce a set of molecular rotational transition lines. If we consider
clouds with cosmological primordial composition, then the only molecule that
contributes to the microwave band with optically thick 
lines is $LiH$, whose lowest rotational
transition occurs at $\nu_0 = 444$ GHz with broadening $\sim 10^{-5}$
(due to the turbulent velocity of molecular clouds in dark clusters). 
This line would be detectable using the Doppler shift effect.
To this aim, it is convenient to consider M31 galaxy, for
whose halo we assume the same picture as outlined above for our 
galaxy. Then we expect that molecular 
clouds should have typical rotational speeds of 50-100 km s$^{-1}$. 
Given the fact that the clouds possess a peculiar velocity
(with respect to the CBR) the 
emitted radiation will be Doppler shifted, with 
$\Delta\nu /\nu_{0}\sim\pm ~10^{-3}$.
However, the precise chemical composition of molecular clouds in the
galactic halo is unknown. Even if the heavy
molecule abundance is very low (as compared with the abundance in
interstellar clouds), many optically thick lines corresponding to the lowest
rotational transitions would show up in the microwave band. In this case, 
it is more convenient to perform broad-band measurements and the Doppler shift 
effect results in an anisotropy in the CBR. 
Since it is difficult to work with fields of view
of a few arcsec, we propose to measure the 
CBR anisotropy between two fields of view - on opposite
sides of M31 - separated by $\sim 4^0$ and with angular resolution
of $\sim 1^0$. We suppose that the halo of M31 consists of
$\sim 10^6$ dark clusters which lie within 25-35 kpc.
Scanning an annulus of $1^0$ width and internal angular
diameter $4^0$, centered at M31, in 180 steps of $1^0$, we would find
anisotropies of $\sim 10^{-5} ~f~ \bar\tau$ in $\Delta T/T$. Here, most of 
the uncertainties arise from the estimate of the
average optical depth $\bar\tau$, which mainly depends
on the molecular cloud composition. In conclusion, since the theory does not
allow to establish whether the expected anisotropy lies above or below
current detectability ($\sim 10^{-6}$), only observations can resolve this
issue.
 
{\bf Infrared searches} -- Another possibility of detecting MACHOs
is via their infrared emission. 
In order to be specific, let us assume that all
MACHOs have same mass 0.08 $M_{\odot}$ and age
$10^{10}$ yr. Accordingly, their surface temperature is
$\sim 1.4 \times 10^3$ K and they emit most of their radiation
(as a black body) at $\nu_{max} \sim 11.5 \times 10^{13}$ Hz.
First, we consider MACHOs  located in M31.
In this case, we find a surface brightness 
$I_{\nu_{max}} \sim 2.1 \times 10^3~(1-f) $ Jy sr$^{-1}$ and
$0.5 \times 10^{3}~(1-f)$ Jy sr$^{-1}$
for projected separations from the M31 center $b=20$ kpc and 40 kpc, 
respectively. Although these values are about one order of magnitude below the 
sensitivity of the detectors on ISO Satellite, they lie above the threshold of 
the detector abord the future planned SIRFT Satellite.
For comparison, we recall that the halo of our galaxy would have in
the direction of the galactic pole a surface brightness
$I_{\nu_{max}}\sim 2 \times 10^{3}~{\rm Jy~sr^{-1}}$, provided 
MACHOs make up the total halo dark matter.
Nevertheless, the infrared radiation
originating from  MACHOs in the halo of our galaxy can be recognized (and
subtracted) by its characteristic angular modulation.
Also, the signal from the
M31 halo can be identified and separated from the galactic background via its
b-modulation. 
Next, we point out that the angular size of dark clusters in the
halo of our galaxy at a distance of $\sim 20$ kpc is $\sim 1.8'$ and the
typical separation among them is
$\sim 14'$. As a result, a characteristic pattern of
bright (with intensity $\sim 3\times 10^{-2}$ Jy at $\nu_{max}$ within
angular size $1.8'$) and dark spots 
should be seen by pointing the detector into
different directions.

\bigskip 
\bigskip

\section{References}
\bigskip

All relevant references can be found in our papers:

F. De Paolis, G. Ingrosso, P. Jetzer and M. Roncadelli,
Phys. Rev Lett. 74, 14 (1995); Astron. Astrophys. 295, 567 (1995); 
Comments on Astrophys. 18, 87 (1995); Int. J. Mod. Phys. D 5, 151 (1996); 
Astron. Astrophys. 329, 74 (1998); Mon. Not. R. Astron. Soc. (1998) 
(to appear); Astrophys. J. (1998) (to appear).

F. De Paolis, G. Ingrosso, P. Jetzer, A. Qadir and M. Roncadelli,
Astr. and Astrophys. 299, 647 (1995).

\end{document}